\documentclass[preprint]{elsarticle}

\usepackage{lineno}

\usepackage[dvipsnames]{xcolor}

\usepackage[utf8]{inputenc} 
\usepackage[T1]{fontenc}    
\usepackage{hyperref}       
\hypersetup{
    colorlinks,
    linkcolor={Red},
    citecolor={Green},
    urlcolor={blue}
}
\usepackage{url}            

\usepackage{amsmath}
\usepackage{amsfonts} 
\usepackage{arydshln}
\usepackage{algorithm}
\usepackage{algpseudocode}
\usepackage{bm,upgreek}
\usepackage{cancel}
\usepackage{caption}
\usepackage{subcaption}
\usepackage{pgfplots}
\pgfplotsset{compat=1.17} 
\usepackage{neuralnetwork}

\newcommand{\p}{\partial}
\newcommand{\R}{\mathbb{R}}
\newcommand\Xcancel[2][black]{\renewcommand\CancelColor{\color{#1}}\xcancel{#2}}
\newcommand{\etal}{\textit{et al}.}

\let\oldequation\equation
\let\oldendequation\endequation

\renewenvironment{equation}
  {\linenomathNonumbers\oldequation}
  {\oldendequation\endlinenomath}

\journal{MASKED}

\begin{document}
\newcommand{\OLHP}[1]{%
\centering
\begin{neuralnetwork}[height=4.5, 
nodespacing=14mm, nodesize=25pt,
layerspacing=23mm]
\fontsize{10}{10}
\newcommand{\x}[2]{$x^{##1}_##2$}
\newcommand{\y}[2]{$\tilde{u}$}
\newcommand{\linklabel}[4]{$w^{2}_##2$}
\inputlayer[count=3, bias=false, title=Input, text=\x]
\hiddenlayer[count=3, bias=false, title=First\\ hidden\\layer, text=\x]
\linklayers
\hiddenlayer[count=4, bias=false, title=Last\\ hidden\\layer, text=\x] 
\linklayers
\outputlayer[count=1, title=Output\\layer, text=\y]

\draw[thick, ->] (7.2,-2.9) -- (8.2,-1.3);
\node[align=center] at (9.1, -1.0) {
Data loss
};

\draw[thick, ->] (7.5,-3.3) -- (8.1,-3.3);
\node[align=center] at (9.1, -4.9) {AutoDiff};
\node[fill=blue!35,  rounded corners=.4cm, align=center, inner sep=.2cm] at (9.1, -3.3) {
$\dfrac{\p\tilde{u}}{\p\mathbf{x}^{0}}$,
\\\\
$\dfrac{\p}{\p \mathbf{x}^{0}}\dfrac{\p\tilde{u}}{\p \mathbf{x}^{0}}$,
\\\\
$\cdots$
};
\draw[thick, ->] (10.1,-3.3) -- (10.7,-3.3);

\ifnum#1=0

\linklayers
\node[align=center] at (11.5, -3.3) {
PDE loss\\
IC loss\\
BC loss
};

\else
\link[from layer=2, from node=1, to layer=3, to node=1, label=\linklabel]
\link[from layer=2, from node=2, to layer=3, to node=1, label=\linklabel]
\link[from layer=2, from node=3, to layer=3, to node=1, label=\linklabel]
\link[from layer=2, from node=4, to layer=3, to node=1, label=\linklabel]
\node[align=center] at (11.5, -3.3) {
\Xcancel[red]{PDE loss}\\
IC loss\\
BC loss
};

\node[align=center] at (2.3,-6.2) 
    {$\underbrace{\quad\quad\quad} $}; 
\node[fill=pink, rounded rectangle, rounded rectangle east arc=none, align=center, anchor=east] at (2.35,-6.9) 
    {$\mathbf{W}^{1}$};
\node[fill=blue!35, rounded rectangle, rounded rectangle west arc=none, align=center, anchor=west] at (2.3,-6.9) 
    {$\,\mathbf{z}^{1}$};
\node[align=center] at (2.3,-7.5) 
    {$\downarrow$};
    
\node[align=center] at (4.6,-6.2) 
    {$\underbrace{\quad\quad\quad} $}; 
\node[fill=pink, rounded rectangle, rounded rectangle east arc=none, align=center, anchor=east] at (4.65,-6.9) 
    {$\mathbf{W}^{2}$};
\node[fill=blue!35, rounded rectangle, rounded rectangle west arc=none, align=center, anchor=west] at (4.6,-6.9) 
    {$\,\mathbf{z}^{2}$};
\node[align=center] at (4.6,-7.5) 
    {$\downarrow$};
    
\node[fill=green!40, rounded rectangle, align=center, inner sep=0.2cm] at (-.1,-8.2) 
    {PDE\\coefficients};
\node[align=center] at (1.2,-8.2) 
    {$\rightarrow$};
\node[draw, align=center, inner sep=.2cm] at (3.5,-8.2) 
    {Steps 1$\sim$5 in Algorithm~1};
\node[align=center] at (5.8,-8.2) 
    {$\rightarrow$}; 

\node[fill=blue!35, rounded rectangle, align=center] at (7.6,-8.2) 
    {$\mathbf{g}^{[1]}, \mathbf{g}^{[2]}, \mathbf{g}^{[3]}
    , \mathbf{g}^{[4]}$}; 

\node[fill=pink, rounded rectangle, align=center, inner sep=.2cm] at (7.8,-6.9) 
    {Trainables \\
    $\lambda^{[2]}, \lambda^{[3]}
    , \lambda^{[4]}$}; 

\node[align=center] at (9.4,-8.1) 
    {$\nearrow$}; 
    
\node[align=center] at (9.4,-7.1) 
    {$\searrow$}; 

\node[fill=orange!50, rounded rectangle,inner sep=.2cm, align=center] at (10.3,-7.6) 
    {$\mathbf{w}^{2}$}; 
    
\draw [->, thick, orange] (10.3,-7.2) to [out=70,in=-70] (5.8,-4.8);    
\fi

\end{neuralnetwork}
}

\begin{frontmatter}



\title{On the Compatibility between Neural Networks and Partial Differential Equations for Physics-informed Learning}


\author[inst1]{Kuangdai Leng}

\affiliation[inst1]{organization={Scientific Computing Department, STFC},
addressline={Rutherford Appleton Laboratory}, 
            city={Didcot},
            postcode={OX11 0QX}, 
            country={UK}}

\author[inst1]{Jeyan Thiyagalingam}

\begin{abstract}
We shed light on a pitfall and an opportunity in physics-informed neural networks (PINNs). We prove that a multilayer perceptron (MLP) only with ReLU (Rectified Linear Unit) or ReLU-like Lipschitz activation functions will always lead to a vanished Hessian. Such a network-imposed constraint contradicts any second- or higher-order partial differential equations (PDEs). Therefore, a ReLU-based MLP cannot form a permissible function space for the approximation of their solutions. Inspired by this pitfall, we prove that a linear PDE up to the $n$-th order can be strictly satisfied by an MLP with $C^n$ activation functions when the weights of its output layer lie on a certain hyperplane, as called the out-layer-hyperplane. An MLP equipped with the out-layer-hyperplane becomes ``physics-enforced'', no longer requiring a loss function for the PDE itself (but only those for the initial and boundary conditions). Such a hyperplane exists not only for MLPs but for any network architecture tailed by a fully-connected hidden layer. To our knowledge, this should be the first PINN architecture that enforces point-wise correctness of PDEs. We show a closed-form expression of the out-layer-hyperplane for second-order linear PDEs, which can be generalised to higher-order nonlinear PDEs.
\end{abstract}

\begin{graphicalabstract}
\resizebox{\textwidth}{!}{\OLHP{1}}
\end{graphicalabstract}

\begin{highlights}
\item We prove that a multilayer perceptron (MLP) with only ReLU-like activation functions will always lead to a vanished Hessian, and thus cannot provide a feasible solution space for any second- or higher-order partial differential equations (PDEs).
\item We prove that an MLP with $C^n$ activation functions can \emph{always} (i.e., regardless of data) satisfy a $n$-th-order linear PDE \emph{exactly} (i.e., zero generalisation error) if the weights of its output layer lie on a hyperplane decided by the given PDE.
\item We give a closed-form expression of this hyperplane for second-order linear PDEs along with an implementation. To the best of our knowledge, this should be the first network architecture that enforces point-wise correctness of PDEs (instead of their initial or boundary conditions).
\end{highlights}

\begin{keyword}
partial differential equation \sep deep learning \sep physics-informed neural network \sep ReLU activation function
\end{keyword}

\end{frontmatter}

\nolinenumbers

\section{Introduction}
\begin{figure}
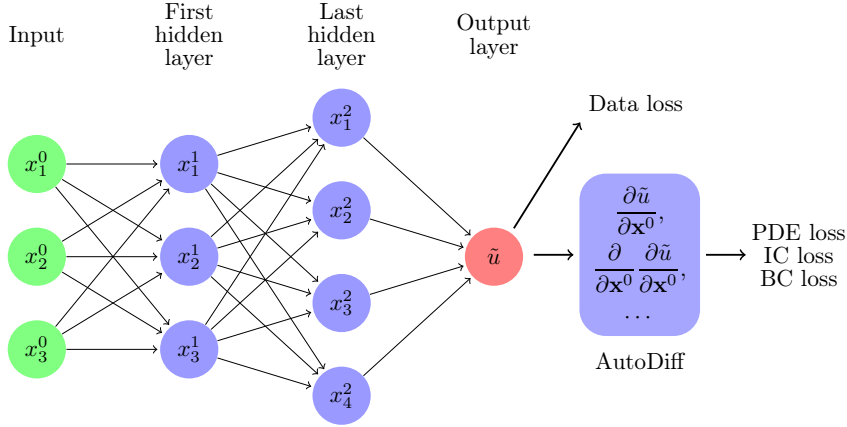

    \centering
    \resizebox{\textwidth}{!}{\OLHP{0}
    }
    \caption{A vanilla PINN with an MLP architecture. 
    The input of the MLP or $\mathbf{x}^{0}$ contains the independent variables of the PDE, and the output, $\tilde{u}(\mathbf{x}^{0})$, is the wanted approximate solution to the PDE.  The acronyms ``IC'' and ``BC'' are used to denote initial and boundary conditions, respectively.
    }
    \label{fig:pinn}
\end{figure}

Simulation and inversion of partial differential equations (PDEs) play a crucial role in applied physics and scientific computing. With conventional numerical methods being computationally expensive, neural networks and deep learning technologies, in particular the physics-informed neural networks (PINNs), have made a new and promising route for simulating and inverting PDEs. Originally formalised by Raissi~\etal~\cite{raissi2019physics}, PINNs have found numerous applications across various domains~\cite{karniadakis2021physics, cuomo2022scientific}, featuring remarkable technical advancements from many aspects, such as the extension to integro-differential equations~\cite{lu2021deepxde}, graph neural operators~\cite{li2020multipole}, identification of nonlinear operators~\cite{lu2021learning}, multi-fidelity models~\cite{meng2020composite}, scalable learning in subdomains~\cite{moseley2021finite} and variational inference~\cite{yang2021b}. Different from data-driven surrogate models~\cite{kasim2021building} aimed only at minimising misfit to data, PINNs are trained also to minimise the physics-based loss functions for the governing equation and the initial and boundary conditions. These PDE-based loss functions are expected to aid the neural networks in understanding the underlying physics whereby to achieve a lower generalisation error with a reduced amount of training data.  A vanilla PINN using a multilayer perceptron (MLP) architecture is shown in Fig.~\ref{fig:pinn}, where the key is to utilise the utmost strength of deep neural networks in automatic differentiation (AutoDiff).

Despite their initial success across many domains, PINNs are still far from being qualified as a routine for solving or inverting PDEs. Many previous studies have reported or been motivated by the underperformance of a vanilla PINN (i.e., an end-to-end MLP) from different perspectives. Examples relevant to our study here include: training in the Fourier space for better consistency between a PINN and a target PDE in terms of their parameterisations of solutions~\cite{li2020fourier, tancik2020fourier}, using multi-scale or multi-frequency neural networks~\cite{CiCP-28-1970, cai2020phase} or domain decomposition techniques~\cite{moseley2021finite, kharazmi2021hp} to enhance the learnability at high frequencies~\cite{ronen2019convergence}, using adaptive activation functions~\cite{jagtap2020locally} or adaptive data sampling~\cite{mao2020physics, YU2022114823} to address unbalanced gradients of the loss functions~\cite{wang2022and, mishra2022estimates},
and explicitly embedding relevant laws or conditions of physics into a network architecture, such as boundary conditions~\cite{sheng2021pfnn, lagari2020systematic, dong2021method}, physical symmetry~\cite{mattheakis2019physical} and invariance~\cite{ling2016reynolds}. These studies more or less boil down to one fundamental question: \emph{given a data regime, to what extent a neural network can be compatible with the target PDE system}.

For clarity, let us consider the following function spaces. Let $\mathcal{F}_{\mathrm{PDE}}$, $\mathcal{F}_{\mathrm{IC}}$, $\mathcal{F}_{\mathrm{BC}}$ be the function spaces that respectively satisfy the governing equation, the initial conditions and the boundary conditions. Given a well-posed PDE system, their intersection $\mathcal{F}_{\mathrm{PDE}}\cap \mathcal{F}_{\mathrm{IC}}\cap\mathcal{F}_{\mathrm{BC}}$ must be a skeleton set, i.e., it must contain one and only one element, as denoted by $u$, which is the true solution. A neural network also form a function space parameterised by its weights, as denoted by $\mathcal{F}_{\mathrm{NN}}$. Clearly, a neural network can be a good candidate for physics-informed learning only if $u\in \mathcal{F}_{\mathrm{NN}}$ within the data regime, of which the necessary conditions are $\mathcal{F}_{\mathrm{NN}}\cap \mathcal{F}_{\mathrm{PDE}}\neq \emptyset$,  $\mathcal{F}_{\mathrm{NN}}\cap \mathcal{F}_{\mathrm{IC}}\neq \emptyset$
and $\mathcal{F}_{\mathrm{NN}}\cap \mathcal{F}_{\mathrm{BC}}\neq \emptyset$. Meanwhile, for a neural network to work efficiently (in terms of data quantity and convergence rate), the following differences, $\mathcal{F}_{\mathrm{NN}}- \mathcal{F}_{\mathrm{PDE}}$, $\mathcal{F}_{\mathrm{NN}}- \mathcal{F}_{\mathrm{IC}}$ and $\mathcal{F}_{\mathrm{NN}}- \mathcal{F}_{\mathrm{BC}}$, should be as small as possible. The above mentioned studies can be well associated with these two purposes; for example,~\cite{CiCP-28-1970, cai2020phase, moseley2021finite, kharazmi2021hp,  jagtap2020locally} can be understood as increasing the expressiveness of $\mathcal{F}_{\mathrm{NN}}$ to capture  localised features in $u$ at high frequencies,~\cite{li2020fourier, tancik2020fourier, mattheakis2019physical, ling2016reynolds} as decreasing  $\mathcal{F}_{\mathrm{NN}}- \mathcal{F}_{\mathrm{PDE}}$ by changing basis functions or imposing energy conservation, and~\cite{sheng2021pfnn, lagari2020systematic, dong2021method} as enforcing $\mathcal{F}_{\mathrm{NN}}- \mathcal{F}_{\mathrm{BC}}=\emptyset$ or $\mathcal{F}_{\mathrm{NN}}\in \mathcal{F}_{\mathrm{BC}}$.

In this paper, we will show a case of incompatibility between $\mathcal{F}_{\mathrm{NN}}$ and $\mathcal{F}_{\mathrm{PDE}}$ ($\mathcal{F}_{\mathrm{NN}}\cap \mathcal{F}_{\mathrm{PDE}}= \emptyset$) and provide a simple and novel approach to enforce full compatibility between them ($\mathcal{F}_{\mathrm{NN}}\in\mathcal{F}_{\mathrm{PDE}}$), both regardless of data. We will prove that an MLP with only  Rectified Linear Unit (ReLU) or ReLU-like Lipschitz activation functions will always cause a vanished Hessian, which contradicts any second- or higher-order PDEs. In other words, any approximate solution yielded by a ReLU-based MLP is non-permissible by such PDEs. While this incompatibility has been explicitly mentioned and avoided in some applications~\cite{markidis2021old, moseley2020solving}, we do see quite a few applications having (probably) falling into this trap. Notably, He~\etal~\cite{he2020relu} has shown that ReLU-based MLPs are formally equivalent to the piecewise linear interpolation used in the finite element methods~\cite{igel2017computational},  implying that a ReLU-based MLP may still be available for second- or higher-order PDEs. We argue that such equivalence is only apparent, and the two methods are essentially different in terms of the locality of function parameterisation; to take advantage of such formal equivalence, certain measures (e.g., learning in subdomains~\cite{ moseley2021finite, kharazmi2021hp}) must be taken to bridge the gap between the two. We will revisit this issue in latter part of this paper.

As inspired by the above ReLU-induced incompatibility, we will prove that a \emph{linear} PDE up to the $n$-th order can be \emph{strictly} satisfied by an MLP with $C^n$ activation functions when the weights of its output layer lie on a hyperplane decided by the PDE, which we refer to as the \emph{out-layer-hyperplane}. Such a hyperplane exists for any network architecture tailed by a fully-connected hidden layer. A PINN equipped with the out-layer-hyperplane becomes ``physics-enforced'', or $\mathcal{F}_{\mathrm{NN}}\in\mathcal{F}_{\mathrm{PDE}}$, no longer requiring a loss function for the governing equation. To the best of our knowledge, this should be the first network architecture that enforces point-wise correctness of the governing equation. We will provide a closed-form expression of the out-layer-hyperplane for second-order linear PDEs approximated by an MLP and provide an implementation. We will also discuss its generalisation to other network architectures and higher-order nonlinear PDEs.

The rest of this paper is organised as follows.  In Section~\ref{sec:hessian}, we derive the Hessian of an MLP-approximated solution, followed by Section~\ref{sec:relu-issues} where we discuss the incompatibility caused by ReLU.  Next, we propose the concept of out-layer-hyperplane in Section~\ref{sec:out-layer-hyperplane}, along with its proof and realisation.  After a brief summary of supportive codes in Section~\ref{sec:code}, we conclude this paper in Section~\ref{sec:conclusion}. As for notations adopted in this paper, we use superscripts to denote layer indices (not powers, except those on $\R$) and subscripts to denote element indices.  Einstein summation notation is assumed for subscript indices except those put in parentheses. We also omit the dot symbol for inner product. For example, a matrix multiplication $\mathbf{A}\cdot\mathbf{B}$ will be written as $\mathbf{A}\mathbf{B}=A_{ij}B_{jk}=\sum_j A_{i(j)}B_{(j)k}$.

\section{Hessian of an MLP approximation}
\label{sec:hessian}

Let $u(\mathbf{x}^{0})\in \R$ denote the true solution of a PDE, where $\mathbf{x}^{0}\in \R^{d^0}$ are the independent variables such as the spatial and temporal coordinates. Taking a 2D homogeneous wave equation on a membrane for example, we have $\mathbf{x}^{0}=\{x, y, t\}$ with $d^0=3$, and $u(\mathbf{x}^{0})$ satisfies the governing equation $u_{xx}+u_{yy}-u_{tt}/c^2=0$ (with $c$ being the wave velocity) and certain initial and boundary conditions. For simplicity, here we assume scalar-valued PDEs, but our conclusions hold for vector-valued ones (e.g., $\mathbf{u}(\mathbf{x}^{0})\in \R^3$).

Let $\tilde{u}(\mathbf{x}^{0})$ be a solution approximated by an MLP that has $L$ layers, the $k$-th layer of which has $d^k$ neurons with weights $\mathbf{W}^{k}\in \R^{d^k\times d^{k-1}}$ and biases $\mathbf{b}^{k}\in \R^{d^k}$.  This approximate solution can be formulated as the following composition of functions:
\begin{equation}
    \tilde{u}(\mathbf{x}^{0})=\mathcal{L}^L\circ
    \sigma^{L-1} \circ \mathcal{L}^{L-1}\circ
    \cdots \circ
    \sigma^{2} \circ \mathcal{L}^{2}\circ
    \sigma^{1} \circ \mathcal{L}^{1}(\mathbf{x}^{0}),
    \label{eq:ufunc}
\end{equation}
where $\mathcal{L}^k$ is the affine transformation endowed by the $k$-th layer: 
\begin{equation}
\mathbf{z}^{k}:=\mathcal{L}^k(\mathbf{x}^{k-1}) = \mathbf{W}^{k} \mathbf{x}^{k-1}+\mathbf{b}^{k},
    \label{eq:zz}
\end{equation}
for $\mathbf{x}^{k-1}\in \R^{d^{k-1}}$ and
    $\mathbf{z}^{k}\in \R^{d^k}$, 
and $\sigma^k$ is the activation function of the $k$-th layer,
\begin{equation}
x^{k}_i=\sigma^k(z^{k}_i),\quad  i\in\{1,2,\cdots,d^k\}.  
\end{equation} 
Because we assume $d^{L}=1$, we write $\mathbf{W}^{L}$ as $\mathbf{w}^{L}$ for a consistent notation.

Based on the chain rule, the Jacobian of $\tilde{u}$, i.e., $\nabla\tilde{u}\in \R^{d^0}$, is given by
\begin{equation}
    \nabla \tilde{u}=
    \mathbf{w}^{L}
    \mathbf{F}^{L-1} \mathbf{W}^{L-1}
    \cdots 
    \mathbf{F}^{2} \mathbf{W}^{2}
    \mathbf{F}^{1} \mathbf{W}^{1},
    \label{eq:nabla}
\end{equation}
where $\mathbf{F}^{k}\in \R^{d^k\times d^k}$ is a diagonal matrix (it is diagonal because $\sigma^k$ is an element-wise operation) containing the first derivatives of $\sigma^k(z)$: 
\begin{equation}
    \mathbf{F}^{k}:=\frac{\p \mathbf{x}^{k}}{\p \mathbf{z}^{k}}=\mathrm{diag}\ \Big\{
    \left.\frac{d\sigma^k}{dz}\right|_{z=z^{k}_1}, \left.\frac{d\sigma^k}{dz}\right|_{z=z^{k}_2}, \cdots, 
    \left.\frac{d\sigma^k}{dz}\right|_{z=z^{k}_3}\Big
    \}.
    \label{eq:F}
\end{equation} 
For the purpose of generalising our results to other network architectures and PDE classes, we rewrite $\nabla\tilde{u}$ in the following top-down manner: 
\begin{equation}
    \nabla \tilde{u}=
    \mathbf{w}^{L}\mathbf{U}=
    \mathbf{w}^{L}
    \mathbf{P}^{k}
    \mathbf{F}^{k} \mathbf{Q}^{k},\quad \forall k\in\{1,2,\cdots,L-1\},
    \label{eq:nablak}
\end{equation}
for which we define
\begin{linenomath}
\begin{align}
    \mathbf{P}^{k}:=&\ 
    \frac{\p \mathbf{x}^{L-1}}{\p \mathbf{x}^{k}}=
    \mathbf{F}^{L-1} \mathbf{W}^{L-1}
    \cdots
    \mathbf{F}^{k+1}\mathbf{W}^{k+1}
    \in \R^{d^{L-1}\times d^{k}},
    \label{eq:P}\\
    \mathbf{Q}^{k}:=&\ 
    \frac{\p \mathbf{z}^{k}}{\p \mathbf{x}^{0}}=
    \mathbf{W}^{k} \mathbf{F}^{k-1} \mathbf{W}^{k-1}
    \cdots
    \mathbf{F}^{2} \mathbf{W}^{2}
    \mathbf{F}^{1} \mathbf{W}^{1}
    \in \R^{d^{k}\times d^{0}},
    \label{eq:Q}\\
    \mathbf{U}:=&\ 
    \frac{\p \mathbf{x}^{L-1}}{\p \mathbf{x}^{0}}=
   \mathbf{P}^{k}
    \mathbf{F}^{k} \mathbf{Q}^{k}\in \R^{d^{L-1}\times d^{0}},\quad \forall k\in\{1,2,\cdots,L-1\}. 
    \label{eq:Udef}
\end{align}
\end{linenomath}

To show the Hessian of $\tilde{u}$, i.e., $\nabla\nabla\tilde{u}\in \R^{d^0\times d^0}$, we first express eq.~\eqref{eq:nablak} in index notation, 
\begin{equation}
    \tilde{u}_{,m}=
    w_i^{L}
    P_{ij}^{k}
    F_{jl}^{k} 
    Q_{lm}^{k},\quad \forall k\in\{1,2,\cdots,L-1\},
    \label{eq:wPFQ}
\end{equation}
where $()_{,i}:=\frac{\p }{\p x_i^{0}}()$. The Hessian of $\tilde{u}$ can be shown by the following steps:
\begin{linenomath}
\begin{subequations}
\begin{align}
     \tilde{u}_{,mn}=
    \frac{\p  \tilde{u}_{,m}}{\p x_n^{0}}=&\ 
     w_i^{L}
     \sum_{k=1}^{L-1}
     P_{ij}^{k}
     \frac{\p F_{jl}^{k}}{\p x^{0}_n} Q_{lm}^{k}\label{eq:line1}
     \\
     =&\ 
     w_i^{L}
     \sum_{k=1}^{L-1}
     P_{ij}^{k}
     \frac{\p F_{jl}^{k}}{\p z^{k}_p}
     \frac{\p z^{k}_p}{\p x^{0}_n}
     Q_{lm}^{k}\label{eq:line2}\\
     =&\ 
     w_i^{L}
     \sum_{k=1}^{L-1}
     P_{ij}^{k}
     \frac{\p F_{jl}^{k}}{\p z^{k}_p}
     Q_{pn}^{k}
     Q_{lm}^{k}
     \label{eq:line3}\\
     =&\ 
     w_i^{L}
     \sum_{k=1}^{L-1}
     \sum_{j=1}^{d^k}
     P_{i(j)}^{k}
     s_{(j)}^{k}
     Q_{(j)m}^{k}
     Q_{(j)n}^{k}.\label{eq:line4}
\end{align}
\label{eq:nabla2}
\end{subequations}
\end{linenomath}
In the above derivation, eq.~\eqref{eq:line1} takes the derivative of \emph{each} $\mathbf{F}^{k}$ in eq.~\eqref{eq:nabla} with respect to $\mathbf{x}^{0}$ and sums them up, showing the result in the spirit of eq.~\eqref{eq:wPFQ};  eq.~\eqref{eq:line2} expands $\frac{\p \mathbf{F}^{k}}{\p \mathbf{x}^{0}}$ by the chain rule, i.e., $\frac{\p \mathbf{F}^{k}}{\p \mathbf{x}^{0}}=\frac{\p \mathbf{F}^{k}}{\p \mathbf{z}^{k}} \frac{\p \mathbf{z}^{k}}{\p \mathbf{x}^{0}}$;  eq.~\eqref{eq:line3} substitutes $\frac{\p \mathbf{z}^{k}}{\p \mathbf{x}^{0}}$ with $\mathbf{Q}^{k}$ as per its definition;  and, finally, eq.~\eqref{eq:line4} takes into account that  $\frac{\p F_{jl}^{k}}{\p z^{k}_p}$ vanishes unless $j=l=p$, as $\sigma^k$ is element-wise, whereas the non-zero terms $s_{j}^{k}$ are the second derivatives of $\sigma^k(z)$:
\begin{equation}
    s_{j}^{k}:=\left.\frac{d}{dz}\frac{d\sigma^k(z)}{dz}\right|_{z=z^{k}_j}.
    \label{eq:s}
\end{equation} 
For its later generalisation, we rewrite eq.~\eqref{eq:nabla2} as
\begin{equation}
     \nabla\nabla \tilde{u}=
    \mathbf{w}^{L}\mathbf{V},
    \label{eq:nabla2k}
\end{equation}
where $\mathbf{V}$ is a third-order tensor defined by
\begin{linenomath}
\begin{subequations}
\begin{align}
\mathbf{V}:=&\ 
    \frac{\p}{\p \mathbf{x}^{0}}\frac{\p \mathbf{x}^{L-1}}{\p \mathbf{x}^{0}}\in \R^{d^{L-1}\times d^{0}\times d^{0}},\label{eq:vline1}\\
     V_{imn}=&\ 
     \sum_{k=1}^{L-1}
     \sum_{j=1}^{d^k}
     P_{i(j)}^{k}
     s_{(j)}^{k}
     Q_{(j)m}^{k}
     Q_{(j)n}^{k}.
     \label{eq:vline2}
\end{align}
\label{eq:Vdef}
\end{subequations}
\end{linenomath}

\section{Incompatibility by ReLU}
\label{sec:relu-issues}

For a $\sigma^k$ of Lipschitz continuity like ReLU, we have $\mathbf{s}^{k}\equiv\mathbf{0}$. It is then straightforward to see from eq.~\eqref{eq:nabla2k} that the Hessian $\nabla\nabla \tilde{u}\equiv\mathbf{0}$ if the $(L-1)$ activation functions are \emph{all} ReLU-like. We provide a simple code to verify this; see Section~\ref{sec:code}. Such a vanished Hessian as imposed by the neural network is unwanted, which limits $\tilde{u}$ to a function space that excludes the true solution $u$, or $\mathcal{F}_\mathrm{NN}\cap\mathcal{F}_\mathrm{PDE}=\emptyset$, regardless of the training data. From the perspective of Probably Approximately Correct (PAC) learning~\cite{shalev2014understanding}, $\mathcal{F}_\mathrm{NN}\cap\mathcal{F}_\mathrm{PDE}=\emptyset$ means the breakdown of the ``realisability assumption'', which fundamentally weakens a model's learnability. For better understanding, let us exam the 2D wave equation on a membrane, considering the following three cases:

\begin{enumerate}
\item
if the wave equation is inhomogeneous, i.e., $u_{xx}+u_{yy}-u_{tt}/c^2=f$ with $f=f(x,y,t)$ being the source term, it can never be satisfied by a ReLU-based $\tilde{u}$ because we have proved that $\tilde{u}_{xx}=\tilde{u}_{yy}=\tilde{u}_{tt}\equiv0$;
\item
if the wave equation is homogeneous, i.e., $u_{xx}+u_{yy}-u_{tt}/c^2=0$, it will always be satisfied by a ReLU-based $\tilde{u}$; however, this illusory benefit is invalid because the network-imposed conditions  (not only $\tilde{u}_{xx}=\tilde{u}_{yy}=\tilde{u}_{tt}\equiv0$ but also $\tilde{u}_{xy}=\tilde{u}_{xt}=\tilde{u}_{yt}\equiv0$) are much stronger than the wave equation itself;
\item
if the wave equation is both forced and damped, i.e, $u_{xx}+u_{yy}-u_{tt}/c^2-\mu u_{t}/c^2=f$ (with $\mu$ being the coefficient of friction), a ReLU-based MLP will effectively be trained to minimise $-\mu\tilde{u}_{t}/c^2-f$, subject to $\nabla\nabla \tilde{u}\equiv\mathbf{0}$, rather than to minimise $\tilde{u}_{xx}+\tilde{u}_{yy}-\tilde{u}_{tt}/c^2-\mu \tilde{u}_{t}/c^2-f$ as wanted.
\end{enumerate}

The vanished Hessian can be easily avoided by using a smooth activation function, as many applications have done, with some being alert to the downside of ReLU for PINNs~\cite{markidis2021old, moseley2020solving}. Another way is to use non-affine layer operations, or non-affine $\mathcal{L}$'s in eq.~\eqref{eq:ufunc}, because $(\sigma \mathcal{L})''$ will contain at least one non-zero term ${\sigma}' \mathcal{L}'$; an example of this is the Fourier neural operators tailed by ReLU~\cite{li2020fourier}. Note that convolutional filters are also affine (with a sparse $\mathbf{W}^{k}$), so a ReLU-based convolutional neural network (CNN) will also end up with a vanished Hessian.

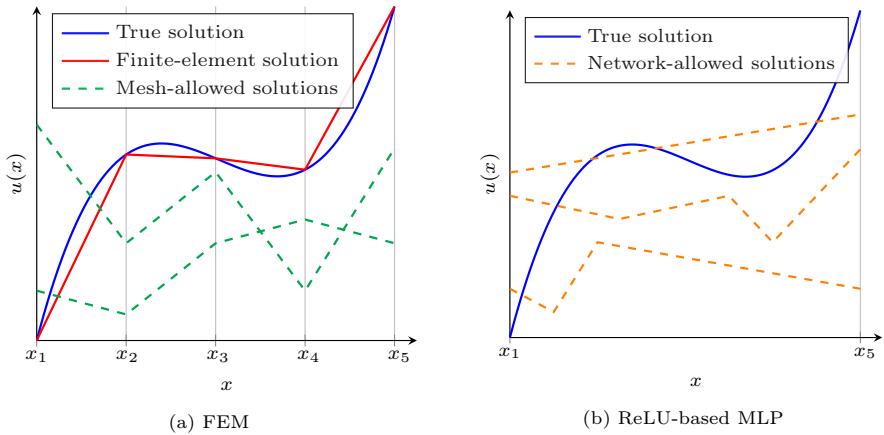
\begin{figure}
\fontsize{7}{6}
    \centering
    \begin{subfigure}{.48\textwidth}
    \resizebox{\textwidth}{!}{
    \begin{tikzpicture}
\begin{axis}[
    axis lines = left,
    xlabel = \(x\),
    ylabel = {\(u(x)\)},
    x label style={at={(.5, -0.1)},anchor=north},
    y label style={at={(-0.1, .5)},anchor=north},
    xticklabels=\empty,
    yticklabels=\empty,
    ymajorticks=false,
    legend style={at={(0.04,.98)},anchor=north west,opacity=.9},
    xtick={-10,-6,-2,2,6},
    xmajorgrids=true,
    legend cell align={left},
    xmax=7,
    width=1.1\textwidth,
        height=\textwidth,
]
\addplot [
thick,
     domain=-10:6, 
    samples=100, 
    color=blue,
]
{.2* x^3 + 1.1*x^2 - 2*x - 1};

\addplot [
thick,
     domain=-10:6, 
    samples=5, 
    color=red,
]
{.2* x^3 + 1.1*x^2 - 2*x - 1};

\addplot [
thick,dashed,
     domain=-10:6, 
    samples=5, 
    color=Green,
]
coordinates {
    (-10,20)(-6,-30)(-2,0)(2,-50)(6,10)
    };

\addplot [
thick,dashed,
     domain=-10:6, 
    samples=5, 
    color=Green,
]
coordinates {
    (-10,-50)(-6,-60)(-2,-30)(2,-20)(6,-30)
    };

\addlegendentry{True solution}
\addlegendentry{Finite-element solution}
\addlegendentry{Mesh-allowed solutions}
\end{axis}
\node at (0,-.2){$x_1$};
\node at (1.15,-.2){$x_2$};
\node at (2.3,-.2){$x_3$};
\node at (3.45,-.2){$x_4$};
\node at (4.6,-.2){$x_5$};
\end{tikzpicture}}
    \caption{FEM}
    \label{fig:linear-fem}
    \end{subfigure}
    \hfill
    \begin{subfigure}{.48\textwidth}
    \resizebox{\textwidth}{!}{
    \begin{tikzpicture}
\begin{axis}[
    axis lines = left,
    xlabel = \(x\),
    ylabel = {\(u(x)\)},
    x label style={at={(.5, -0.1)},anchor=north},
    y label style={at={(-0.1, .5)},anchor=north},
    xticklabels=\empty,
    yticklabels=\empty,
    ymajorticks=false,
    legend style={at={(0.04,.98)},anchor=north west,opacity=.9},
    xtick={-10,6},
    xmajorgrids=true,
    legend cell align={left},
    xmax=7,
    width=1.1\textwidth,
        height=\textwidth,
]
\addplot [
thick,
     domain=-10:6, 
    samples=100, 
    color=blue,
]
{.2* x^3 + 1.1*x^2 - 2*x - 1};

\addplot [
thick,dashed,
     domain=-10:6, 
    samples=5, 
    color=orange,
]
coordinates {
    (-10,-10)(-5,-20)(-0,-10)(2,-30)(6,10)
    };

\addplot [
thick,dashed,
     domain=-10:6, 
    samples=5, 
    color=orange,
]
coordinates {
    (-10,-50)(-8,-60)(-6,-30)(6,-50)
    };
    
\addplot [
thick,dashed,
     domain=-10:6, 
    samples=5, 
    color=orange,
]
coordinates {
    (-10,-0)(6,25)
    };

\addlegendentry{True solution}
\addlegendentry{Network-allowed solutions}
\end{axis}
\node at (0,-.2){$x_1$};
\node at (4.6,-.2){$x_5$};
\end{tikzpicture}
    }
    \caption{ReLU-based MLP}
    \label{fig:linear-mlp}
    \end{subfigure}
    \caption{Sketch of 1D piecewise linear parameterisations (a) by an FEM and (b) by a ReLU-based MLP.
    The main difference between (a) and (b) is that the anchor points are predefined in (a), i.e., $x_1,x_2,\cdots,x_5$, based on the required mesh resolution, but are stochastic and uncontrollable in (b).     A simple code is provided to produce the sketched patterns in (b) with randomly initialised MLPs; see Section~\ref{sec:code}.
    }
    \label{fig:linear}
\end{figure}

He~\etal~\cite{he2020relu} have studied the similarity between ReLU-based MLPs and the finite element methods (FEMs) with a linear shape function~\cite{igel2017computational}. They show that both of them present a piecewise linear function space for the approximation of the PDE solution, which seems to hint that a ReLU-based MLP may achieve the same accuracy as a linear FEM does. We argue that such a similarity or equivalence is only apparent because of the following two major differences. First, in an FEM, the spatial domain is discretised by a mesh whereby linear interpolation is localised in each (small) element with fixed anchor points (i.e., the nodes). Such a local parameterisation with compact support can fit any complicated functions within in the mesh resolution. In contrast, a ReLU-based MLP presents a global piecewise linear parameterisation whose anchor points are stochastically located, due to the stochastic nature of neural weights, mostly uncontrollable a priori and difficult to learn from data. Such  a difference is visualised in Fig.~\ref{fig:linear}. 
Second, an FEM is based on a \emph{weak form} (or variational form) of the PDE~\cite{kharazmi2021hp} in which the second-order derivatives have disappeared.
However, such a weak form is not considered by a vanilla PINN, which still computes the Hessian by AutoDiff as part of the loss function  (and it will only get zeros if ReLU is used). The above two differences are most relevant to the idea of PINNs with domain decomposition~\cite{moseley2021finite, kharazmi2021hp}, that is, many smaller neural networks are trained in parallel, each approximating the solution in a smaller subdomain, either using a weak form~\cite{kharazmi2021hp} or a strong form~\cite{moseley2021finite} of the PDE. Domain decomposition can be understood as to achieve controlled localisation of basis functions by using many neural networks, similar to an FEM by using many elements. In addition to the weak form, an alternative way to avoid computing the Hessian is to reformulate a second-order PDE as a system of first-order PDEs~\cite{gladstone2022fo}, similar to the staggered-grid finite difference methods~\cite{igel2017computational}.

\section{The out-layer-hyperplane}
\label{sec:out-layer-hyperplane}

In the previous section, we have shown that a ReLU-based MLP is incompatible with a second-order PDE, or $\mathcal{F}_\mathrm{NN}\cap\mathcal{F}_\mathrm{PDE}=\emptyset$. The above process has delightfully inspired a simple and elegant way to enforce an MLP to satisfy any second-order linear PDEs, i.e., $\mathcal{F}_\mathrm{NN}\in\mathcal{F}_\mathrm{PDE}$, without imposing any non-physical constraints. Now we assume that the activation functions are of $C^2$ or above so that $\mathbf{V}$ in eq.~\eqref{eq:Vdef} does not vanish.

Consider a general second-order, scalar-valued linear PDE as follow:
\begin{equation}
    \Gamma_{mn} u_{,mn}+\gamma_{m} u_{,m}+\beta u+\alpha=0,
    \label{eq:PDE}
\end{equation}
where $\alpha \in \R$, $\beta \in \R$, $\bm{\upgamma} \in \R^{d^0}$ and $\mathbf{\Gamma} \in \R^{d^0\times d^0}$ are known PDE coefficients.  Let $u(\mathbf{x}^{0})$ be approximated by $\tilde{u}(\mathbf{x}^{0})$.  Substituting $u$ by $z^L$ in eq.~\eqref{eq:zz}, $u_{,m}$ by eq.~\eqref{eq:nablak} and $u_{,mn}$ by eq.~\eqref{eq:nabla2k}, we obtain the following linear system that enforces the MLP to satisfy the above PDE:
\begin{equation}
    \psi_i w_i^{L}+\beta b^{L}+\alpha=0,
    \label{eq:wb}
\end{equation}
where $\bm{\uppsi}\in\R^{d^{L-1}}$ are defined by
\begin{equation}
    \psi_i:=\Gamma_{mn} V_{imn}
    +\gamma_mU_{im}
    +\beta x_i^{L-1}.
    \label{eq:psi}
\end{equation} The LHS of eq.~\eqref{eq:wb} is a function of all the weights and biases of the MLP and the input $\mathbf{x}^{0}$. However, from the angle of network design, we can regard the weights and biases of the hidden layers ($\mathbf{w}^{k}$ and $b^{k}$, $k\in\{1,2,\cdots,L-1\}$) as free variables or trainable parameters while constraining those of the output layer ($\mathbf{w}^{L}$ and $b^{L}$) so that eq.~\eqref{eq:wb} can always hold. Obviously, $\mathbf{w}^{L}\in\R^{d^{L-1}}$ and $b^{L}\in\R$ lie on a hyperplane in $\R^{d^{L-1}+1}$, as defined by eq.~\eqref{eq:wb}, which we call the \emph{out-layer-hyperplane}.

\begin{figure}
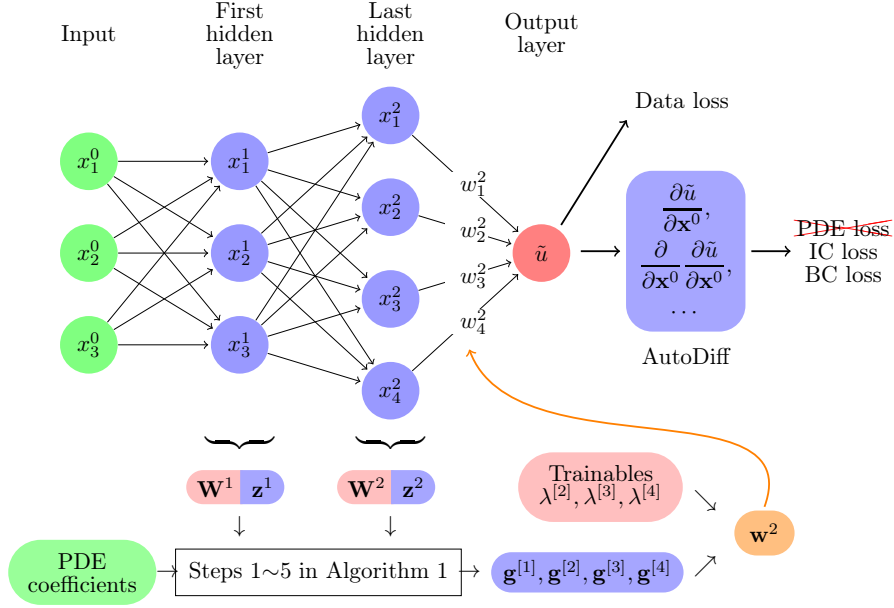

    \centering
    \resizebox{\textwidth}{!}{\OLHP{1}}
    \caption{Forward pass through an MLP ($L=3$) equipped with the out-layer-hyperplane.     In an ordinary MLP, the four elements of $\mathbf{w}^{2}$ (coloured in orange) are trained.     Using the out-layer-hyperplane, the three $\lambda$'s (coloured in pink) become the new trainable parameters, which always yield a $\mathbf{w}^{2}$ lying on the out-layer-hyperplane so that $\tilde{u}$ always satisfies the target linear PDE. 
    }
    \label{fig:OLHP}
\end{figure}

We realise the out-layer-hyperplane using its parametric equation. Without loss of generality, we ignore $b^{L}$ by assuming $\beta=0$ for a compact notation. Known from basic linear algebra, the parametric equation of the hyperplane is given by
\begin{equation}
    \mathbf{w}^{L}=\mathbf{g}^{[1]}+\sum_{p=2}^{d^{L-1}} \lambda^{[p]}  \mathbf{g}^{[p]},
    \label{eq:wL}
\end{equation}
where the basis vectors $\mathbf{g}^{[p]}\in \R^{d^{L-1}}$, $p\in\{1,2,\cdots,d^{L-1}\}$, can be chosen as 
\begin{equation}
\renewcommand{\arraystretch}{1.3}
\left[
\begin{array}{c}
    \mathbf{g}^{[1]} \\\hdashline[2pt/2pt]
    \mathbf{g}^{[2]} \\\mathbf{g}^{[3]}
    \\\vdots\\ \mathbf{g}^{[d^{L-1}]}
\end{array}
\right]
=
\left[
\begin{array}{c;{2pt/2pt}cccccc}
  \frac{-\alpha \psi_1}{|\bm{\uppsi}|^2}&
  \frac{-\alpha \psi_2}{|\bm{\uppsi}|^2}&
  \frac{-\alpha \psi_3}{|\bm{\uppsi}|^2}
  &\cdots &
  \frac{-\alpha \psi_{d^{L-1}}}{|\bm{\uppsi}|^2}
  \\\hdashline[2pt/2pt]
  -\psi_2  & \psi_1 &0&\cdots&0\\
  -\psi_3  &0&\psi_1&\cdots&0\\
  \vdots&\vdots&\vdots&\ddots&\vdots\\ -\psi_{d^{L-1}}&0&0&\cdots&\psi_1
\end{array}
\right],
\label{eq:v}
\end{equation}
and $\lambda^{[p]}\in \R$, $p\in\{2,3,\cdots,d^{L-1}\}$ (note that $p$ starts from 2 here), are the new network parameters taking the place of $\mathbf{w}^{L}$.  Here we put the superscript indices for $\lambda$ and $\mathbf{g}$ in $[\cdot]$ to differentiate them with the layer indices. Note that $\mathbf{g}^{[1]}$ represents a point on the hyperplane, and $\mathbf{g}^{[2]}, \mathbf{g}^{[3]}, \cdots, \mathbf{g}^{[d^{L-1}]}$ are linearly independent vectors parallel to the hyperplane. The choice of these basis vectors is non-unique, such as $\mathbf{g}^{[1]}\leftarrow\{-\alpha/\psi_1, 0, 0, \cdots, 0\}$, but our choice in eq.~\eqref{eq:v} avoids division by elements of $\bm{\uppsi}$ (which may be close to zero),  while the division by $|\bm{\uppsi}|^2$ is always safe and stable. Also note that the total number of the $\lambda$'s is one less than the size of $\mathbf{w}^{L}$, as $\mathbf{w}^{L}$ must lie on a hyperplane, which certifies that we do not impose any extra conditions other than the PDE itself. The forward pass through an MLP equipped with the out-layer-hyperplane is sketched in Fig.~\ref{fig:OLHP} and elaborated in Algorithm~\ref{alg:forward}. A code implementation is also provided; see Section~\ref{sec:code}.

Thus far, we have shown the existence and a realisation of the out-layer-hyperplane for second-order linear PDEs approximated by an MLP. What makes this idea more attractive is that it can be generalised to other network architectures and PDE classes, most of which require a trivial effort, as detailed below.
\begin{itemize}
    \item \textbf{Network architecture.} 
    Equation~\eqref{eq:wb} holds for any network architecture tailed by a fully-connected hidden layer as long as we relax the definitions of $\mathbf{U}$ and $\mathbf{V}$ in eqs.~\eqref{eq:Udef} and \eqref{eq:Vdef} respectively to $\mathbf{U}:= 
    \frac{\p \mathbf{x}^{L-1}}{\p \mathbf{x}^{0}}$ and 
    $\mathbf{V}:= 
    \frac{\p}{\p \mathbf{x}^{0}}
    \frac{\p \mathbf{x}^{L-1}}{\p \mathbf{x}^{0}}$ (i.e., abandon eq.~\eqref{eq:vline2}).     For an MLP, we have given their closed-form expressions in eqs.~\eqref{eq:Udef} and \eqref{eq:Vdef}, corresponding to Steps 2$\sim$4 in Algorithm~\ref{alg:forward}.     For a general network architecture, we can simply compute $\mathbf{U}$ and $\mathbf{V}$ by AutoDiff using the first $(L-1)$ layers based on their generalised definitions.

    \item \textbf{Higher-order linear PDEs.} 
    It is straightforward to show that the out-layer-hyperplane exists for an $n$-th order linear PDE given $C^n$ activation functions.      For example, a general third-order linear PDE will require an extra term $\Omega_{mnr} u_{,mnr}$ in eq.~\eqref{eq:PDE}, with $\Omega_{mnr}$ being known PDE coefficients;
    this term will add $\Omega_{mnr} T_{imnr}$ to the hyperplane coefficients ${\psi_i}$ in eq.~\eqref{eq:psi},  where $\mathbf{T}:= 
    \frac{\p}{\p \mathbf{x}^{0}}
    \frac{\p}{\p \mathbf{x}^{0}}
    \frac{\p \mathbf{x}^{L-1}}{\p \mathbf{x}^{0}}$, and $\mathbf{T}\neq\mathbf{0}$ with $C^3$ or above activation functions.

    \item \textbf{Vector-valued linear PDEs.} 
    Let us consider a general vector-valued, second-order linear PDE with solution $\mathbf{u}(\mathbf{x}^{0})\in\R^3$,
    formulated as
    \begin{equation}
        \Theta_{ijmn} u_{j,mn}+\Pi_{ijm} u_{j,m}+\Lambda_{ij} u_j+\alpha_i=0,\quad i\in \{1,2,3\},
    \end{equation}
    which is the $\R^3$ version of eq.~\eqref{eq:PDE}.     Consequently, one can show that eq.~\eqref{eq:wb} will be generalised to
    \begin{equation}
        \Psi_{ijk} W_{jk}^{L} + \Lambda_{ij} b^{L}_j +\alpha_i =0,\quad i\in \{1,2,3\},
    \end{equation}
    where $\Psi_{ijk}=\Theta_{ijmn}V_{kmn}+\Pi_{ijm}U_{km}+\Lambda_{ij} x_k^{L-1}$, with $\mathbf{U}$ and $\mathbf{V}$ remaining the same as in eqs.~\eqref{eq:Udef} and \eqref{eq:Vdef}.     Evidently, the above linear system states that the PDE can be enforced if the weights and biases of the output layer, $\mathbf{W}^{L}\in\R^{3\times d^{L-1}}$ and $\mathbf{b}^{L}\in\R^3$, lie on three coupled hyperplanes, which can be realised by parametric equations similar to eq.~\eqref{eq:wL}.

    \item\textbf{Nonlinear PDEs.}
    It is not difficult to guess that,  to satisfy a nonlinear PDE, the weights of the output layer must lie on a hypersurface. This is true.      For example, consider $(u_{,i} u)$, a common nonlinear term that appears in many PDEs such as Navier–Stokes and Bateman-Burgers~\cite{debnath2005nonlinear}.      It will introduce many quadratic terms about $\mathbf{w}^{L}$ into eq.~\eqref{eq:wb}, namely, $A_{ijk} w_{j}^{L} w_{k}^{L}$, where $A_{ijk}= U_{ij} x_{k}^{L-1}$.     For a general nonlinear PDE, there is no universal way to parameterise its out-layer-hypersurface.     For certain PDEs, however, a closed-form parameterisation similar to eq.~\eqref{eq:wL} may be found.     At least, concerning second- and third-order nonlinear PDEs, the parametric equations for the generalised quadratic and cubic hypersurfaces have been well established~\cite{bajaj1989quadric}.  
\end{itemize}

\begin{algorithm}
\caption{Forward pass through an MLP equipped with the out-layer-hyperplane.}
\label{alg:forward}
\textbf{Input}: $\mathbf{x}^{0}\in \R^{d^0}$.\\
\textbf{Output}: $\tilde{u}\in \R$.\\
\textbf{Network parameters} (for backprop): 
$\mathbf{W}^{k}\in \R^{d^{k}\times d^{k-1}}$ and $\mathbf{b}^{k}\in \R^{d^{k}}$, 
$k\in\{1,2,\cdots,L-1\}$; 
$\lambda^{[p]}\in\R$, $p\in\{2,3,\cdots,d^{L-1}\}$.\\
\textbf{PDE coefficients}:
$\alpha \in \R$, $\beta \in \R$, $\bm{\upgamma} \in \R^{d^0}$ and $\mathbf{\Gamma} \in \R^{d^0\times d^0}$.
\vspace{.1cm}
\begin{algorithmic}[1]
\State Perform ordinary forward pass until the output layer, obtaining $\mathbf{x}^{L-1}\in\R^{d^{L-1}}$ and $\mathbf{z}^{k}\in \R^{d^{k}}$, 
$k\in\{1,2,\cdots,L-1\}$; {\color{OliveGreen}\# $\mathbf{z}^{k}$'s are usually unsaved in an ordinary MLP, but we need them to compute the first and second derivatives of the activation functions;}
\State Compute $\mathbf{F}^{k}\in \R^{d^k \times d^k}$ and $\mathbf{s}^{k}\in \R^{d^k}$, $k\in\{1,2,\cdots,L-1\}$ by eqs.~\eqref{eq:F} and \eqref{eq:s};
\State Compute $\mathbf{P}^{k}\in\R^{d^{L-1}\times d^k}$ and $\mathbf{Q}^{k}\in \R^{d^k\times d^0}$ $k\in\{1,2,\cdots,L-1\}$ by eqs.~\eqref{eq:P} and \eqref{eq:Q};
\State Compute $\mathbf{U}\in\R^{d^{L-1}\times d^0}$ and $\mathbf{V}\in\R^{d^{L-1}\times d^0\times d^0}$ by eqs.~\eqref{eq:Udef} and \eqref{eq:Vdef};
\State Compute $\bm{\uppsi}\in\R^{d^{L-1}}$ by eq.\eqref{eq:psi} and then $\mathbf{g}^{[p]}\in \R^{d^{L-1}}$, $p\in\{1,2,\cdots,d^{L-1}\}$ by \eqref{eq:v}; {\color{OliveGreen}\# this is where the PDE coefficients are used;}
\State Reconstruct $\mathbf{w}^{L}\in \R^{d^{L-1}}$ by \eqref{eq:wL}; {\color{OliveGreen} \# this is where the network parameters $\lambda^{[p]}$'s are used;}
\State Return $\tilde{u}=\mathbf{w}^{L}\mathbf{x}^{L-1}+b^{L}$.
\end{algorithmic}

\end{algorithm}

\newpage

\section{Validation by code}
\label{sec:code}
To support some of the key statements in this paper,
we provide a lightweight code repository (\url{https://github.com/stfc-sciml/PINN-PDE-compatibility}) from which the following three standalone PyTorch~\cite{paszke2019pytorch} scripts can be found:
\begin{itemize}
    \item \texttt{relu\_causes\_zero\_hessian.py} verifies that a ReLU-based MLP will always lead to a vanished Hessian;
    \item \texttt{piecewise\_linear\_by\_relu.py} plots the piecewise linear functions generated by some random ReLU-based MLPs, as a support to our sketch in Figure~\ref{fig:linear-mlp};
    \item \texttt{out\_layer\_hyperplane.py} implements the out-layer-hyperplane in an MLP for a 2D inhomogeneous wave equation on a membrane:
    $u_{xx} + u_{yy} - u_{tt }- u_t=\sin(x + y - t)$, following Algorithm~\ref{alg:forward}. 
\end{itemize}

\section{Conclusions}
\label{sec:conclusion}

In physics-informed learning, it is crucial that the function space granted by a neural network, $\mathcal{F}_\mathrm{NN}$, is compatible with that required by the target PDE, $\mathcal{F}_\mathrm{PDE}$. We have proved that an MLP with only ReLU-like activation functions will always lead to a vanished Hessian, so it is incompatible with any second- or higher-order PDEs, namely, $\mathcal{F}_\mathrm{NN}\cap\mathcal{F}_\mathrm{PDE}=\emptyset$. Therefore, we proclaim that ReLU should not be used in a vanilla PINN. This has led us to a simple method to make $\mathcal{F}_\mathrm{NN}$ fully compatible with $\mathcal{F}_\mathrm{PDE}$, namely, $\mathcal{F}_\mathrm{NN}\in\mathcal{F}_\mathrm{PDE}$. We have proved that, using sufficiently smooth activation functions, a neural network tailed by a fully-connected hidden layer can strictly satisfy any high-order linear PDE when the weights of its output layer lie on a certain hyperplane, as called the out-layer-hyperplane. A closed-form expression of this hyperplane has been given for second-order linear PDEs. To the best of our knowledge, this should be the first PINN architecture that enforces point-wise correctness of  PDEs. Our future work will be focused on the efficacy and performance of the out-layer-hyperplane for different types of architectures and PDEs.

\section*{Acknowledgement}
\noindent This work is supported by the EPSRC grant, Blueprinting for AI for Science at Exascale (BASE-II, EP/X019918/1), which is Phase~II of the Benchmarking for AI for Science at Exascale (BASE) grant.

 \bibliographystyle{elsarticle-num} 
 \bibliography{cas-refs}





\end{document}